\documentclass[aip,apl,graphicx,preprint]{revtex4-2}

\usepackage{amsmath, amsfonts, amstext, amssymb}
\usepackage[english]{babel}
\usepackage{subfigure}
\usepackage[thinlines]{easytable}
\usepackage{siunitx}
\usepackage{exscale}
\usepackage{tabulary}
\usepackage{ae}
\usepackage{graphicx, wrapfig}
\usepackage{enumerate}
\usepackage{wrapfig}
\usepackage{placeins}
\usepackage[dvipsnames]{xcolor}

\begin{document}

\title{Ultrafast coherent THz lattice dynamics coupled to spins in a van der Waals antiferromagnetic flake}%

\author{F. Mertens}
\affiliation{Department of Physics, TU Dortmund University,
	Otto-Hahn Stra\ss{}e 4, 44227 Dortmund, Germany}
\author{D. M\"{o}nkeb\"{u}scher}
\affiliation{Department of Physics, TU Dortmund University,
	Otto-Hahn Stra\ss{}e 4, 44227 Dortmund, Germany}
\author{E. Coronado}
\affiliation{Instituto de Ciencia Molecular (ICMol) Universidad de Valencia. Catedrático José Beltrán 2 46890, Paterna, Spain}
\author{S. Ma\~{n}as-Valero}
\affiliation{Instituto de Ciencia Molecular (ICMol) Universidad de Valencia. Catedrático José Beltrán 2 46890, Paterna, Spain}
\author{C. Boix-Constant}
\affiliation{Instituto de Ciencia Molecular (ICMol) Universidad de Valencia. Catedrático José Beltrán 2 46890, Paterna, Spain}
\author{A. Bonanni}
\affiliation{Institute of Semiconductor and Solid State Physics, Johannes Kepler University Linz, Altenbergerstr. 69, 4040 Linz, Austria}
\author{M. Matzer}
\affiliation{Institute of Semiconductor and Solid State Physics, Johannes Kepler University Linz, Altenbergerstr. 69, 4040 Linz, Austria}
\author{R. Adhikari}
\affiliation{Institute of Semiconductor and Solid State Physics, Johannes Kepler University Linz, Altenbergerstr. 69, 4040 Linz, Austria}
\author{A. M. Kalashnikova}
\affiliation{Ioffe Institue, 194021 St. Petersburg, Russia}
\author{D. Bossini} \email[]{davide.bossini@uni-konstanz.de}
\affiliation{Department of Physics and Center for Applied Photonics, University of Konstanz, D-78457 Konstanz, Germany.}
\affiliation{Department of Physics, TU Dortmund University,
	Otto-Hahn Stra\ss{}e 4, 44227 Dortmund, Germany}
\author{M. Cinchetti}
\affiliation{Department of Physics, TU Dortmund University,
	Otto-Hahn Stra\ss{}e 4, 44227 Dortmund, Germany}
\date{\today}%

 \begin{abstract}
A coherent THz optical lattice mode is triggered by femtosecond laser pulses in the antiferromagnetic van der Waals semiconductor FePS$_3$. The 380 nm thick exfoliated flake was placed on a substrate and laser-driven lattice and spin dynamics were investigated as a function of the excitation photon energy and sample temperature. The pump-probe spectroscopic measurements reveal that the photo-induced phonon is generated by a displacive mechanism. The amplitude of the phononic signal decreases as the sample is heated up to the N\'eel temperature and vanishes as the phase transition to the paramagnetic phase occurs. This evidence confirms that the excited lattice mode is intimately connected to the long-range magnetic order. Therefore our work discloses a pathway towards a femtosecond coherent manipulation of the magneto-crystalline anisotropy in a van der Waals antiferromagnet. In fact, it is reported that by applying a magnetic field the induced phonon mode hybridizes via the Kittel-mechanism with zone-centre magnons.
%
 \end{abstract}


\maketitle

\pagenumbering{arabic}

The synthesis of few-atomic-layers-thin materials\cite{burch2018, park2016, gibertini_2019} has ignited the spark of a massive research effort aiming at manipulating their macroscopic properties. More recently almost-two-dimensional (2D) magnetically ordered materials have been produced as well\cite{huang2017, gong2017}. The long-range magnetic order in these compounds appears to be highly susceptible to lattice distortions. In particular, static strains were shown to play a decisive role in emergence of ferromagnetism in the 2D limit of CrI$_3$, which is antiferromagnetic in a bulk form \cite{thiel2019}. Strains affect also the magnetic order in the MPS$_3$ (M=Ni,Fe,Mn) material systems\cite{chittari2016, ni2021}. These effects are rooted in the role of the magnetic anisotropy in the stabilisation of the long-range order in 2D magnets\cite{mermin1966}, as the microscopic origin of the magneto-crystalline anisotropy can be expressed in term of the spin-orbit coupling $\lambda \bold{L} \cdot\bold{S}$, where $\lambda$ is the coupling constant, while $\bold{L}$ and $\bold{S}$ represent the orbital and spin momentum\cite{stohr_book}. Recently, the ultrafast  generation of phonons, via a variety of mechanisms, has been demonstrated to be a powerful tool for driving and controlling spin dynamics in bulk magnets at fundamental timescales \cite{nova2016, disa2020, juraschek2020, stupakiewicz2021, formisano2021, soumah2021}. Optically  driven collective lattice mode carry therefore tremendous potential for the manipulation of the long-range magnetic order in 2D magnets, in particular considering the well-established possibility to drive such modes fully coherently even with a photon-energy far from their eigenfrequency\cite{Merlin1997,dresselhaus1992}. In this work we experimentally investigate a flake of the van der Waals antiferromagnet FePS$_3$, demonstrating the coherent laser-induced excitation of an optical lattice mode, intimately coupled to the long-range magnetic order, since it can be detected only below the ordering temperature and, additionally, it is reported to hybridize with a zone-centre magnon mode, under the application of an external field\cite{zhang_coupling_2021, vaclavkova2021, liu2021}.

\begin{figure}
\includegraphics[width=\textwidth]{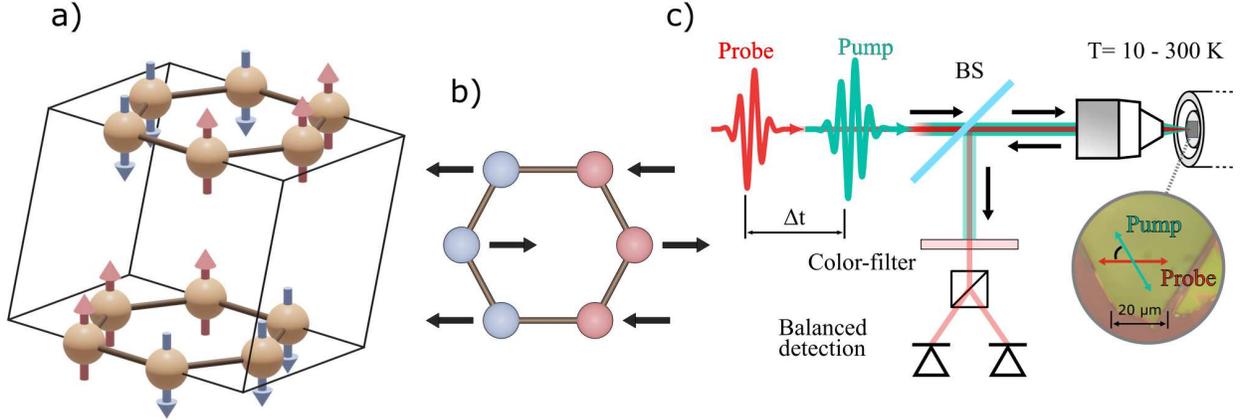}
\caption{\footnotesize{(a) Fe-ions within the magnetic unit cell of FePS3. (b) Raman-active phonon mode with $\omega$=3.2\,THz observed in FePS$_3$ \cite{zhang_coupling_2021}. (c) Schematic view of the experimental setup. The pump (0.83\,eV - 1.08\,eV) and probe (1.45\,eV) beams are collinear and focussed on the same surface using a 50x magnification, infinity corrected, apochromatic microscope objective. The beams reflected by the sample surface are reflected by a beam splitter (BS) to the balanced detection scheme. The pump is filtered out from the detector employing a colour filter.}}
\label{fig:setup}
\end{figure}

An interesting class of van der Waals antiferromagnets is represented by the MPS$_3$ compounds\cite{chittari2016, coak2019,lee2016, kim2019} (M=Ni,Fe,Mn). While a coherent optical generation of magnons has been reported in a single-crystal free standing (i.e. no substrate) bulk NiPS$_3$\cite{afanasiev2021}, this material is not promising in terms of scaling the concept to the 2D limit. In fact it has been experimentally demonstrated that a single atomic layer of NiPS$_3$ is not magnetically ordered\cite{kim2019}, differently from MnPS$_3$\cite{long2020} and FePS$_3$\cite{lee2016}. Therefore we select FePS$_3$ and we investigate a specimen consisting of an exfoliated flake with lateral size of $\approx 50$ $\mu$m deposited on a SiO$_2$/Si substrate. Our sample is not as thin as a single-layer (thickness $\approx 380$ nm, as reported in the Supplementary material), nevertheless it represents a structure that can be scaled down (i.e. exfoliated flake, substrate), in a material already experimentally proven to be antiferromagnetically ordered even in the monolayer limit\cite{lee2016, wang_2016}. Below the N\'eel temperature (T$_N$$\approx$118\,K\cite{lee2016, mccreary2020}) the magnetic moments of the Fe-ions are oriented out-of-plane and form a zigzag pattern within the layers, which build up to a monoclinic crystal structure with a $C2/m$ spacegroup\cite{lancon2016} (see Fig. \ref{fig:setup}(a)). One of the allowed lattice vibrations in this space group\cite{zhang_coupling_2021}, which is relevant for this paper is shown in Figure \ref{fig:setup} (b). The band-gap energy of FePS$_3$ is reported at $\approx 1.5$\,eV\cite{brec1979}, while a $d-d$ transition around 1.1 eV appears in the absorption spectrum measured at 80 K\cite{piacentini1982}.

For our experiment we employ the pump-probe technique to measure the photo-induced rotation of the probe polarisation. The literature abundantly demonstrates that this experimental scheme is able to track the lattice dynamics\cite{Merlin1997}, as well as the complete coherent and incoherent longitudinal and transverse dynamics of the N\'eel vector (which is the antiferromagnetic order parameter)\cite{satoh2010, kampfrath2010, nemec2018, bossini2017, bossini2016, bossini2019}, provided that the symmetry of the investigated material allows quadratic magneto-optical effects, as it is the experimentally proven for FePS$_3$\cite{zhang2021}. Our set-up consists of an amplified laser system with a repetition rate of 200 KHz. The main output of the laser (20\,W average output power) is split into two beams (13\,W and 7\,W) seeding two optical parametric amplifiers (OPA), which generate laser beams with photon energy in the 0.5 -3.5 eV range. The outputs of the OPAs, whose photon energies can be tuned independently of each other, are employed as pump and probe beams, as described elsewhere\cite{mertens2020}. The sample is mounted on a piezo-driven three-axis-stage inside a cryostat (see Fig. \ref{fig:setup}(c)). Moreover both pump and probe beams are focussed on the flake with an objective down to a spot of 1.5 $\mu$m diameter, as estimated by knife-edge measurements. This configuration allows to fully focus both beams into spots on the different parts of the flake, so that homogeneous regions can be addressed\cite{Adhikari2021}. In the experiments, the probe photon-energy was kept constant at 1.45 eV, which is just below the band-gap energy of 1.5\,eV\cite{brec1979}, while the pump photon energy was tuned in a range of 0.83\,eV - 1.08\,eV below the band gap of the substrate. The pump photon energy is not tuned to values higher than 1.1\,eV, as excitation of electrons in the conduction band of Si would than contribute to the signal.

\begin{figure}
\includegraphics{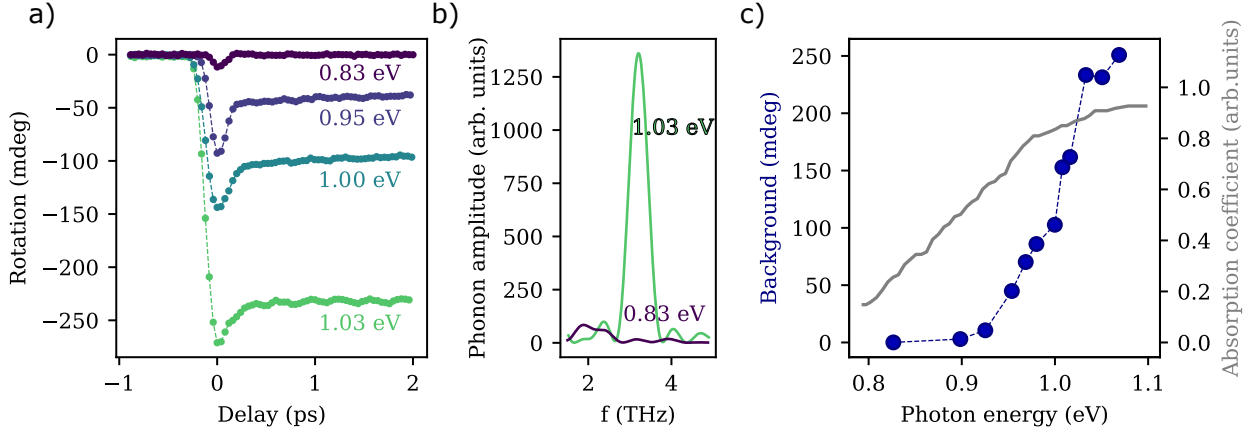}
\caption{\footnotesize{(a) Spectral dependence of the rotation of the polarisation. The sample temperature was set to 10 K and the pump fluence kept constant to 2 mJ/cm$^2$. Pump and probe beam were linearly polarized 60\,$\deg$ away from each other. The photon energy of the pump was tuned. (b) Fourier transformation of the time-traces measured by exciting the sample with pump photon-energies equal to 0.95 eV and 1.03 eV. (c) Contribution of the incoherent background as a function of the pump photon-energy (blue) and the absorption coefficient at 80 K from the literature\cite{piacentini1982} (grey). The error bars lie within the markers.}}
\label{fig:WL_traces}
\end{figure}

Figure \ref{fig:WL_traces}(a) reports the time-resolved data obtained by setting the sample temperature to 10 K. The data display both a coherent and incoherent contribution, strongly dependent on the excitation photon-energy. Some traces reveal coherent oscillations, whose frequency amounts to 3.2 THz, as determined by means of the Fourier transform of the signal (see Fig. \ref{fig:WL_traces}(b)). This value matches the eigenfrequency of the Raman-active phonon shown in Fig. \ref{fig:setup}(b) and reported in the literature\cite{lee2016, wang_2016, mccreary2020,ghosh2021}. Therefore, the coherent oscillations can be readily ascribed to changes of linear crystallographic birefringence by a laser-driven phonon mode. The data shown in Fig. \ref{fig:WL_traces}(a) are just a few selected traces of the set measured as a function of the pump photon-energy. Taking into account all the measurements, it is possible to quantify the spectral dependence of both the incoherent background and the coherent oscillations. In Fig.\ref{fig:WL_traces}(c) we plot the trend of the incoherent background (extrapolated as explained in Supplementary materials) as a function of the excitation photon-energy. In the same figure we show the absorption spectrum of FePS$_3$ reported in the literature\cite{piacentini1982}, measured at T = 80 K. The trend of the background is consistent with the absorption spectrum, suggesting that the microscopic processes originating the incoherent component of the signal are dissipative. The nature of such processes will be discussed later on.

\begin{figure}
\includegraphics[width=0.5\textwidth]{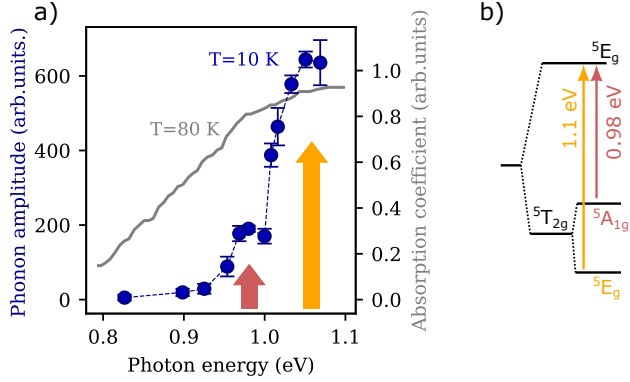}
\caption{(a) Amplitude of the 3.2 THz spectral component (blue markers). The grey line corresponds to the absorption of FePS$_3$ at 80 K, as reported in the literature\cite{piacentini1982}. (b) Schematic of electronic transitions between the crystal field $3d$-state of the Fe$^{2+}$ ions split by the octahedral and trigonal ligand fields\cite{joy1992}.}
\label{fig:WL}
\end{figure}

Similarly, we visualise the spectral dependence of the phonon amplitude in Fig. \ref{fig:WL}(a). This trend was obtained by Fourier transforming each data set and then considering the area of the peak centered at 3.2 THz frequency (described in the Supplementary materials). The amplitude of the phonon resembles the trend of the absorption spectrum, steadily increasing up to 1.1 eV. We note that a peak at 0.98 eV appears, while it is not observed in the absorption spectrum. The absorption spectrum in this region is dominated by crystal field $d-d$ transitions of the Fe$^{2+}$ ions\cite{joy1992}. In particular, these processes have been identified with the two transitions at 0.98\,eV and 1.1\,eV as depicted in Fig. \ref{fig:WL}(b)\cite{joy1992}. The trend reported on Fig. \ref{fig:WL}(a) shows two features, consistent with the presence of two resonances, while the absorption spectrum exhibits a monotonically increasing profile. It is reasonable to suggest that the thermal broadening of the electronic processes is the reason why the spectral line at 0.98 eV, whose fingerprint is detected in our data taken at T = 10 K, does not appear in the absorption spectrum measured at T = 80 K. Considering the Raman-active character of the excited phonon and the correlation between its amplitude and the $d-d$ transitions, we conclude that laser pulses drive the collective lattice excitations by means of the displacive mechanism\cite{dresselhaus1992, Merlin1997}. In a nutshell, this mechanisms relies on the excitation of electrons from an occupied to an unoccupied state. Such process alters the inter-ionic potentials so that the previous atomic positions are no longer equilibrium ones. Hence the atoms start moving towards the minima of the modified potential, at the frequency determined by the dispersion relation of phonons. The linear dependence of the amplitude of the coherent oscillations on the pump fluence further confirms the assignment of the excitation mechanism (see Supplementary Figure 2).

\begin{figure}
\includegraphics{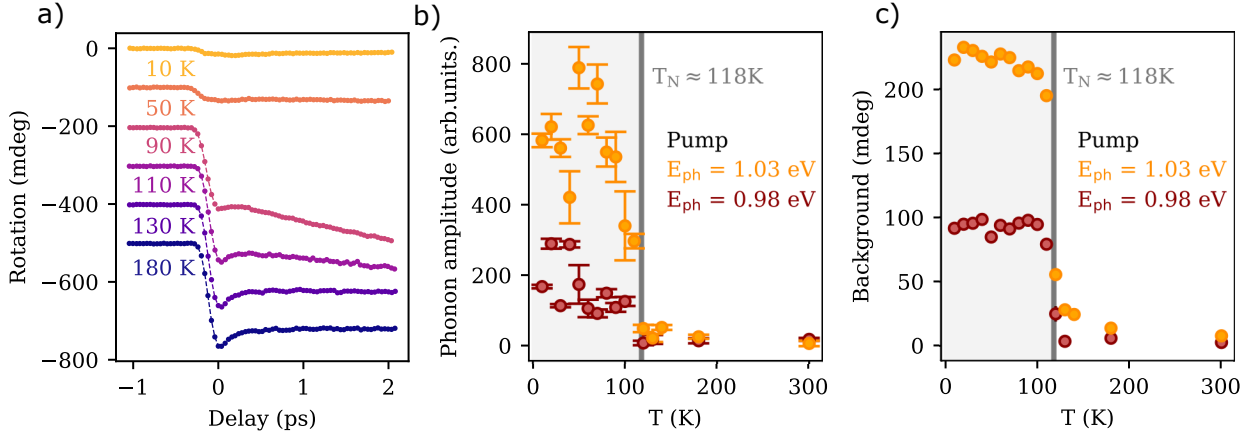}
\caption{\footnotesize{(a) Pump-induced rotation of the probe polarisation detected at different values of tempearture. The excitation photon-energy was set to 1.03 eV. (b) Temperature dependence of the phonon amplitude for excitation photon-energies of 1.03\,eV and 0.98\,eV. (c) Incoherent background contribution as a function of the temperature for excitation photon-energies of 1.03 eV and 0.98 eV. The error bars lie within the markers.}}
\label{fig:T_traces}
\end{figure}


Seeking to experimentally establish a connection between the coherent phonon and the long-range magnetic order, we measure the temperature dependence of the pump-induced rotation of the probe polarisation. The time-traces (see Fig. \ref{fig:T_traces}(a)), obtained by setting the pump photon energy to 1.03 eV, reveal a pronounced  dependence of both the phonon amplitude and the background signal on the sample temperature. Two complete data sets, corresponding to different excitation energies (1.03 eV and 0.98 eV) are processed and analysed. The results are shown in Fig. \ref{fig:T_traces}(b) and show that the amplitude of the phonons vanishes as the N\'eel temperature is approached and crossed. The 3.2 THz phonons can thus be induced and detected only in the presence of the long-range antiferromagnetic order. We note that this is fully consistent with the symmetry of the lattice mode, which is a zone-folded mode. This lattice collective excitations appear in the dispersion of the material only in the antiferromagnetic phase, as the paramagnetic to antiferromagnetic phase transition takes place in concomitance to a doubling of the unit cell, and thus halving of the Brillouin zone\cite{mccreary2020}.

Let us now turn the discussion to the identification of the incoherent background. Analysing the amplitude of the background as a function of temperature (see Fig. \ref{fig:T_traces}(c)), it is straightforward to assess that the background is a feature apparent only in the magnetically-ordered phase. From the data shown in Fig. \ref{fig:WL_traces} we have already concluded that the incoherent contribution to the signal represents dissipative processes. The temperature dependence reveals that the characteristic time associated with the background increases as the N\'eel temperature is approached. We observe that in the literature a critical slow-down of photo-induced incoherent spin dynamics in FePS$_3$ has been already reported\cite{zhang2021}. The literature reports  similar features in the rotation of the probe polarisation, which is interpreted as demagnetisation of the sublattices, provided that quadratic magneto-optical effects are allowed in the used experimental geometry\cite{bossini2014, zhang2021,saidl2017}. Summing up all these considerations and noting a correlation between the background signal and absorption spectrum (Fig. \ref{fig:WL_traces}(c)), we ascribe the incoherent background to the demagnetisation of the two Fe$^{2+}$ sublattices, triggered by photon absorption and dissipative processes\cite{kirilyuk2010,bossini2017}.

In conclusion, we have demonstrated THz coherent lattice and incoherent spin dynamics in the antiferromagnetic phase of FePS$_3$, driven by femtosecond laser pulses in a region of weak absorption. The coherent displacive excitation of a 3.2 THz optical phonon mode, which is intimately connected with the long-range magnetic order, is demonstrated. An incoherent contribution to the signal was observed and identified as the demagnetisation of the sublattices. Considering that our experiment was carried on a flake of FePS$_3$ deposited on a substrate, our results can be scaled down to a coherent THz modulation of the magneto-crystalline anisotropy in thinner flakes, even in the 2D limit, as FePS$_3$ and MnPS$_3$, differently from NiPS$_3$, retains the antiferromagnetic order\cite{kim2019}. In view of the phonon hybridisation with zone centre magnons\cite{zhang_coupling_2021, vaclavkova2021, liu2021}, applying an external magnetic field to this material can result in photoinducing even coherent phonomagnonic THz dynamics in a 2D antiferromagnet.

See the Supplementary material for further description of the data analyis, the linear dependence of the oscillation ampltiude to the pump-fluence and an estimation of the sample thickness.

{\footnotesize{This work was supported by the Deutsche Forschungsgemeinschaft through the International Collaborative Research Centre 160 (Projects B9 and Z3) and by BO 5074/1-1, by the COST Action MAGNETOFON (grant number CA17123). A.B. acknowledges Austrian Science Fund (FWF), Project P31423. A.M.K. acknowledges RFBF (Grant No. 19-52-12065).}

\section*{Data availability}
\normalsize{
The data that support the findings of this study are available from the corresponding author upon reasonable request.}

\bibliography{bibliography}
\end{document}


\title{Ultrafast coherent THz lattice dynamics coupled to spins in a van der Waals antiferromagnetic flake}%

\author{F. Mertens}
\affiliation{Department of Physics, TU Dortmund University,
	Otto-Hahn Stra\ss{}e 4, 44227 Dortmund, Germany}
\author{D. M\"{o}nkeb\"{u}scher}
\affiliation{Department of Physics, TU Dortmund University,
	Otto-Hahn Stra\ss{}e 4, 44227 Dortmund, Germany}
\author{E. Coronado}
\affiliation{Instituto de Ciencia Molecular (ICMol) Universidad de Valencia. Catedrático José Beltrán 2 46890, Paterna, Spain}
\author{S. Ma\~{n}as-Valero}
\affiliation{Instituto de Ciencia Molecular (ICMol) Universidad de Valencia. Catedrático José Beltrán 2 46890, Paterna, Spain}
\author{C. Boix-Constant}
\affiliation{Instituto de Ciencia Molecular (ICMol) Universidad de Valencia. Catedrático José Beltrán 2 46890, Paterna, Spain}
\author{A. Bonanni}
\affiliation{Institute of Semiconductor and Solid State Physics, Johannes Kepler University Linz, Altenbergerstr. 69, 4040 Linz, Austria}
\author{M. Matzer}
\affiliation{Institute of Semiconductor and Solid State Physics, Johannes Kepler University Linz, Altenbergerstr. 69, 4040 Linz, Austria}
\author{R. Adhikari}
\affiliation{Institute of Semiconductor and Solid State Physics, Johannes Kepler University Linz, Altenbergerstr. 69, 4040 Linz, Austria}
\author{A. M. Kalashnikova}
\affiliation{Ioffe Institue, 194021 St. Petersburg, Russia}
\author{D. Bossini} \email[]{davide.bossini@uni-konstanz.de}
\affiliation{Department of Physics and Center for Applied Photonics, University of Konstanz, D-78457 Konstanz, Germany.}
\affiliation{Department of Physics, TU Dortmund University,
	Otto-Hahn Stra\ss{}e 4, 44227 Dortmund, Germany}
\author{M. Cinchetti}
\affiliation{Department of Physics, TU Dortmund University,
	Otto-Hahn Stra\ss{}e 4, 44227 Dortmund, Germany}
\date{\today}%

\maketitle

\section{Data analysis}
\begin{figure}[!htb]
\includegraphics[width=\textwidth]{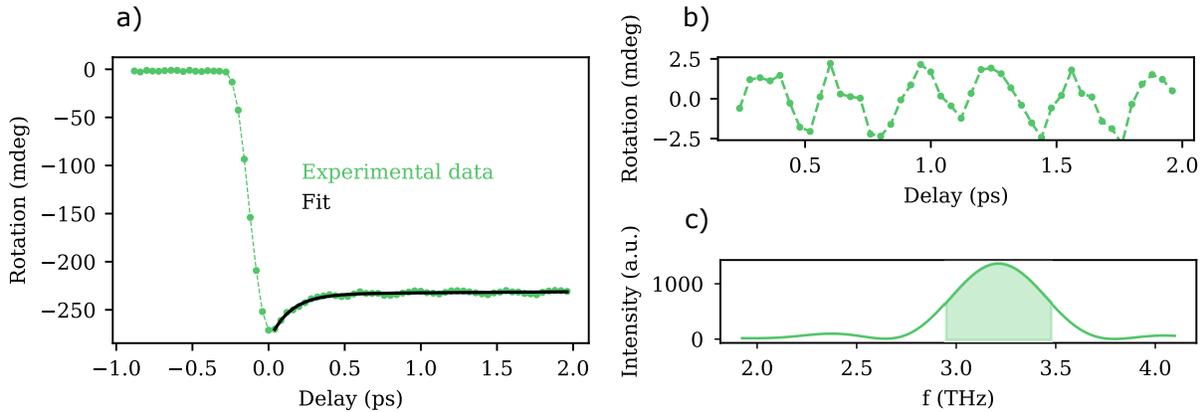}
\caption{a) Rotation of the polarisation induced by a 1.03\,eV pump beam at T=10\,K. The best fit to the data, obtained with Eq. \eqref{eq:fit} is illustrated with the black line. b) The 3.2\,THz oscillation isolated from the incoherent background. c) Square modulus of the Fourier Transform transform of the trace shown in b). The area within the full width at half maximum of the phonon peak is marked in light green.}
\label{fig:fit}
\end{figure}

We describe the data processing aimed at the quantitative evaluation of both the incoherent and coherent contributions to the detected rotation of the probe polarisation.
We fit the signal at positive delays with the following function:
\begin{equation}
f(x)=A_\text{exp}\mathrm{e}^{-t/\tau}+A_\text{lin} t + c
\label{eq:fit}
\end{equation}
We perform a fit to the data from Fig. 2 a) and Fig. 4a) of the main text with Eq. \eqref{eq:fit}.
The best fit to the data was subtracted from the entire time-trace, so that the 3.2 THz phononic oscillation was isolated.
We quantify the amplitude of the lattice mode by Fourier transforming (square modulus of the FFT algorithm) the residual of the data, i.e. the isolated oscillations. We then evaluate the area within the full width at half maximum of the 3.2 THz line, representing the photo-induced phonon.
Supplementary Figure \ref{fig:fit} presents the three data analysis steps for an exemplary pump-probe trace.
We estimate the error of the Fourier transformed data by calculating the standard deviation of a section of the Fourier spectrum which does not show any peak.
The fitting parameter $c$ is taken as a measure for the incoherent background (Fig. 2(c) and Fig. 4(c) in the main text).

\section{Pump-fluence dependence}

\begin{figure}
\includegraphics[width=\textwidth]{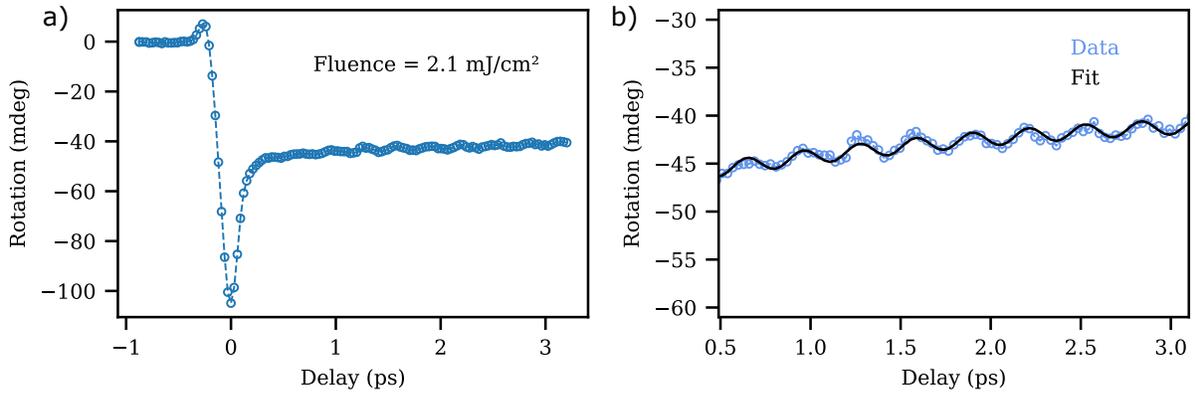}
\caption{(a) Pump-probe trace at a fluence of 2.1\,mJ/cm$^2$. (b) Close-in view of the 3.2\,THz oscillations and the fit function from Eq. \eqref{eq:fit_fluence}.}
\label{fig:fluence_fit}
\end{figure}

\begin{figure}
\includegraphics{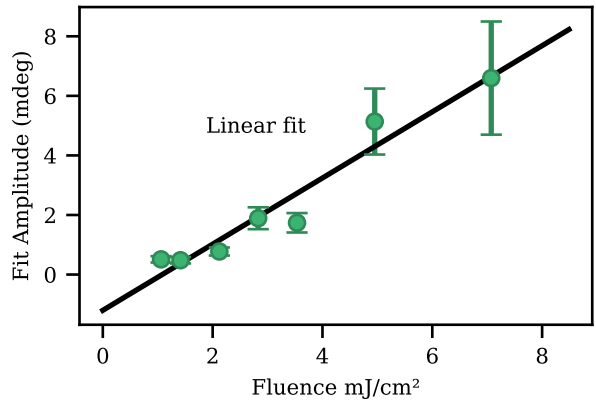}
\caption{Fluence dependence of the phonon amplitude at a pump photon-energy of $E_\text{ph}$ = 1.03\,eV and temperatures of $T$ = 10\,K. The solid line represents a linear regression.}
\label{fig:fluence_amplitude}
\end{figure}

In Supplementary Fig. 2 a) the rotation of the polarisation of an exemplary data set, taken from a fluence dependence measured at 10\,K tuning the pump photon energy to 1.03\,eV, is shown.
We fit the pump-probe traces of this dependence, as displayed in figure \ref{fig:fluence_fit}, with the following function:
\begin{equation}
	h(x)=A\sin(\omega t+\phi) \mathrm{e}^{-t/\tau}+at^2+bt+c
\label{eq:fit_fluence}
\end{equation}
The amplitude $A$ shows a linear trend in the dependence of the phonon oscillation amplitude on the laser fluence (S. Fig. \ref{fig:fluence_amplitude}), which is consistent with the excitation of coherent phonons via a displacive mechanism.
For the calculation of the fluence we measured the spot size by a knife-edge method, using the straight edge of a gold marker on the substrate and by moving it via the piezo driven stage across the beam in the focal position.

\section{Flake thickness}

\begin{figure}
\includegraphics{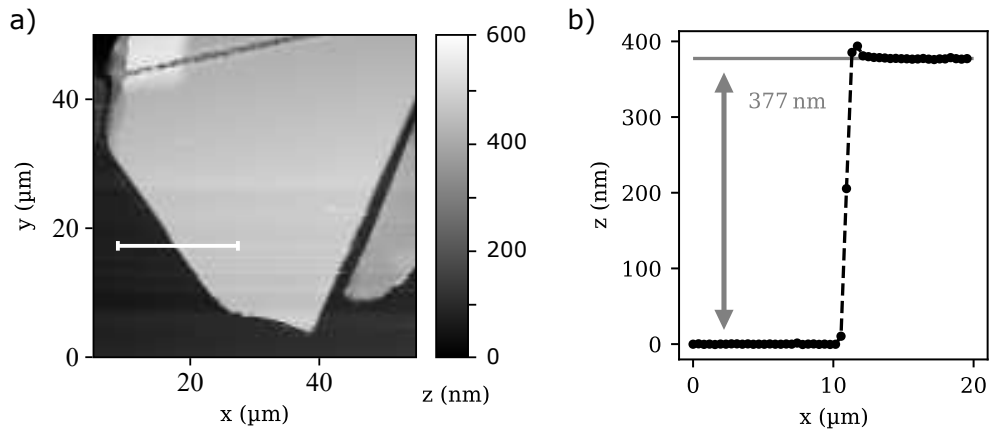}
\caption{(a) AFM image of the investigated flake. (b) Height profile between the flake und substrate surface across the white marker in (a).}
\label{fig:afm}
\end{figure}

We perform atomic force microscopy (AFM) on the sample surface and calculate the thickness of the flake by taking the height difference across the edge between the flake and the substrate (See S. Fig. \ref{fig:afm}).
In this way a step height of 377\,nm was obtained.

